\documentclass[12pt]{article}
\usepackage{amssymb}
\usepackage{epsfig}

\newcommand{\beqn}{\begin{eqnarray}}
\newcommand{\eeqn}{\end{eqnarray}}
\newcommand{\beq}{\begin{equation}}
\newcommand{\eeq}{\end{equation}}
\newcommand{\eq}[1]{(\ref{#1})}
\newcommand{\abstracts}[1]{{\centering{\begin{minipage}{13.0truecm}
 \normalsize\baselineskip=15pt \centerline{\footnotesize
 ABSTRACT}\vspace*{0.3cm} \parindent=20pt #1 \end{minipage}}\par}}
\newcommand{\cA}{{\cal A}}
\newcommand{\GeV}{~\mathrm{GeV}}
\newcommand{\MeV}{~\mathrm{MeV}}

\newcommand{\offdiag}{\mathrm{offdiag}}
\newcommand{\diag}{\mathrm{diag}}

\begin{document}
~ \vspace{-1cm}
\begin{flushright}
{\large ITEP-LAT/2002-20}

\vspace{0.2cm}

{\large KANAZAWA-02-30}

\vspace{0.3cm}
{\sl February 10, 2003}

{\small\sl revised March 9, 2003}
\end{flushright}

\begin{center}
{\baselineskip=24pt {\Large \bf Abelian dominance and gluon propagators in the
Maximally Abelian gauge of $\mathbf{SU(2)}$ lattice gauge theory}\\

\vspace{1cm}

{\large
V.G.~Bornyakov$^{a,b,c}$, M.N.~Chernodub$^{a,b}$,\\ F.V.~Gubarev$^{a}$,
S.M.~Morozov$^{a}$, M.I.~Polikarpov$^{a}$
}
}
\vspace{.5cm}
{\baselineskip=16pt
{ \it

$^{\rm a}$ Institute of Theoretical and  Experimental Physics,
    Moscow, 117259, Russia\\
$^{\rm b}$ Institute for Theoretical Physics, Kanazawa University,
    Kanazawa 920-1192, Japan\\
$^{\rm c}$ Institute for High Energy Physics, Protvino 142284,
    Russia\\
}
}
\end{center}

\vspace{5mm}

\abstracts{Propagators of the diagonal and the off-diagonal gluons are studied
numerically in the Maximal Abelian gauge of $SU(2)$ lattice gauge theory.
It is found that in the infrared region the propagator of the diagonal gluon is
strongly enhanced in comparison with the off--diagonal one.
The enhancement factor is about 50 at our smallest momentum 325~MeV.
We have also applied various fits to the propagator formfactors.}

\date{\today}

\section{Introduction}

\indent
The propagators of the fundamental  fields play
important role in the understanding of the physical structure of any
quantum field theory.
The gluon propagator in QCD is well known in perturbation theory, {\it i.e.} at large momenta.
On the other hand
its form in the infrared region has not been fixed so far although it has been
intensively studied both analytically and numerically using lattice regularization.
Analytical results range from the infrared divergent \cite{QCD:anal-diver} to
infrared vanishing \cite{QCD:anal-vanish} propagator (see also recent review
\cite{Review:propagators}). The recent lattice investigations
\cite{fit2,Nakamura:mp,fit12,Bonnet,Nakajima:2002kh}
excluded the infrared divergent behavior leaving open the possibility of the infrared
vanishing propagator. Another recent study \cite{Alexandrou:2001fh}
-- made in the Laplacian gauge which is free of Gribov copies --
provided some support to the form of the propagator with dynamically generated
gauge invariant mass proposed in \cite{Cornwall:1981zr}.

It is widely believed that the knowledge of
the infrared behavior of the gluon propagator
is crucial for understanding of the confinement problem.
At present there are two competing scenarios
of confinement: condensation of monopoles~\cite{tHooftMandelstam} or
center vortices~\cite{Zelenotochkin}.
The Maximally Abelian (MA) gauge is
the most convenient for demonstration of the dual superconductor nature of the
gluodynamics vacuum (see, {\it e.g.},~\cite{Review} for a review).
The first study of the MA gauge gluon propagator in the coordinate space was made
in~\cite{Amemiya:zf}. It was found that the propagator of the off-diagonal gluons
is exponentially suppressed at large distances by the effective mass about $1.2$~GeV.
Thus the findings of~\cite{Amemiya:zf} support the Abelian dominance in
gluodynamics~\cite{Ezawa,Suzuki:1989gp}. The mass gap generation for the
off-diagonal gluons was further studied analytically in
Ref.~\cite{Schaden:1999ew, Offdiagonal:mass}.

In this paper we consider the propagators in the momentum space which
allows a detailed investigation of their infrared properties
compared to the coordinate space. Our preliminary results were published in~\cite{Bornyakov:2002vv}.
We describe the gauge fixing and present definition of propagators
in Sections~\ref{sec:GaugeFixing} and~\ref{sec:Gribov}.
Section~\ref{sec:Numerical} is devoted to discussion of numerical
results and the last Section contains our conclusions. In Appendix we
discuss the quality of the gauge fixing procedure, Gribov copies and
finite volume effects.

\section{Gauge Fixing}
\label{sec:GaugeFixing}

We use the standard parameterization of $SU(2)$ link matrices
$U_{11}=\cos\varphi\, e^{i\theta}$ and $U_{12}=\sin\varphi\, e^{i\chi}$.
The gauge fields are defined as follows
$$
\frac{1}{2} A^a_\mu(x) \,\sigma^a = \frac{1}{2i}(U_\mu(x) - U_\mu^{\dagger}(x))\,,
$$
where $\sigma^a$ are the Pauli matrices.
In terms of the link angles one gets\footnote{Note that in Ref.~\cite{Bornyakov:2002vv}
the definition of the field $A$ differs from Eq.~\eq{eq:fields} by the factor of $2$.}:
\beqn
   \frac{1}{2} A^1_\mu(x) & = & \sin\varphi_\mu(x)\sin\chi_\mu(x)\,,\nonumber\\
   \frac{1}{2} A^2_\mu(x) & = & \sin\varphi_\mu(x)\cos\chi_\mu(x)\,,
\label{eq:fields} \\
   \frac{1}{2} A^3_\mu(x) & = & \cos\varphi_\mu(x)\sin\theta_\mu(x)\,. \nonumber
\eeqn
We call $A_\mu^3(x)$ the diagonal gluon field, and
$A_\mu^i(x),\ i = 1,2$, the off-diagonal gluon field.

The MA gauge condition in a differential form is~\cite{tHooft}:
\beqn
\Bigl[\partial_\mu \mp i A^3_\mu(x)\Bigr]\, A^{\pm}_\mu(x)=0\,,\quad
A^{\pm}_\mu=\frac{1}{\sqrt{2}}(A^1_\mu \pm i A^2_\mu)\,.
\label{eq:MAG:local}
\eeqn
Note that here and below we are using the same notations for
lattice and continuum fields. A nonperturbative fixing of this
gauge amounts to the minimization of the functional
$$
F^{\mathrm{cont}}_{\mathrm{MAG}}[A] = \int d^4 x \, \Bigl\{[A^1_\mu(x)]^2
+ [A^2_\mu(x)]^2\Bigr\}\,,
$$
which has the following lattice counterpart:
\beqn
F^{\mathrm{latt}}_{\mathrm{MAG}}[A] = \sum_{x,\mu}
\cos2\varphi_\mu(x)\,.
\label{eq:MAG:global}
\eeqn

The MA gauge condition~\eq{eq:MAG:local} leaves $U(1)$ degrees of freedom
unfixed. To complete the gauge fixing we use a $U(1)$ Landau gauge. In
continuum the Landau gauge condition is
\beq
\label{eq:contland}
\partial_\mu A^3_\mu(x) = 0\,.
\eeq
In previous lattice studies \cite{Amemiya:zf,Bornyakov:2002vv}
to fix $U(1)$ Landau gauge
the following lattice functional
\beqn
\label{eq:coscosfunc:U1}
F^{\mathrm{latt}}_{\mathrm{Land}}[\theta, \varphi]
=\sum_{x,\mu} \cos\theta_\mu(x)\,,
\eeqn
was maximized with respect to $U(1)$ gauge transformations, $\theta_\mu(x) \to \theta_\mu(x)
+ \partial_\mu\omega(x)$.

In this paper we implement the following generalization of the $U(1)$
gauge fixing functional~\eq{eq:coscosfunc:U1}
\beq
\label{eq:coscosfunc}
{\tilde F}^{\mathrm{latt}}_{\mathrm{Land}}[\theta, \varphi]=\sum_{x,\mu}
\cos\varphi_\mu(x) \cos\theta_\mu(x)\,,
\eeq
which is consistent with the definition of $A^3_\mu$ in ~\eq{eq:fields}.
Contrary to the definition in
 Eq.\eq{eq:coscosfunc:U1} this condition implies that
$A_\mu^3$
is transverse for any lattice spacing. In the continuum  limit
definitions \eq{eq:coscosfunc:U1} and \eq{eq:coscosfunc}
coincide with each other.

\section{Propagators}
\label{sec:Gribov}

We calculate the diagonal propagator
\beq
\label{eq:diagprop}
D_{\mu\nu}^{\mathrm{diag}}(p) =\langle\cA^3_\mu(k)\cA^3_\nu(-k)\rangle\,,
\eeq
and the off-diagonal propagator
\beq
\label{eq:offdiagprop}
D_{\mu\nu}^{\mathrm{offdiag}}(p)=
\langle\cA^{+}_\mu(k)\cA^{-}_\nu(-k)\rangle,
\eeq
where the Fourier transformed field, $\cA_\mu^i(k)$, is defined as
follows:
\beqn
\cA_\mu^i(k)=\frac{1}{\sqrt{L^4}}
\sum_x e^{-ik_\nu x_\nu-\frac{i}{2}k_\mu}A_\mu^i(x)\,,\quad
k_\mu=\frac{2\pi n_\mu}{aL_\mu}\,, \quad n_\mu = 0,...,L_\mu-1\,.
\eeqn
The standard variables are
$$
p_\mu=\frac{2}{a}\sin{\frac{a k_\mu}{2}}\,,
$$
in terms of which lattice propagator of a free massive scalar particle
in momentum space has a familiar
form, $D(p) \propto 1 \slash {(p^2+m^2)}$.
Moreover, in the lattice momentum space the gauge condition \eq{eq:contland} becomes:
\beq
\label{eq:lg-mom}
p_\mu \cA_\mu^3 = 0 .
\eeq
The most general structure of both diagonal and off-diagonal propagators is
\beq
\label{eq:gendmunu}
D_{\mu\nu}(p) = \Bigl(\delta_{\mu\nu} - \frac{p_\mu p_\nu}{p^2}\Bigr)\, D_t(p^2)
+ \frac{p_\mu p_\nu}{p^2}D_l(p^2)\,,
\eeq
where $D_{t,l}$ are the scalar functions. They are related to the
{\it formfactors} $D_{\mu\nu}(p)$ as follows:
\beqn
\label{eq:structure:functions}
D_l(p^2) = \frac{p_\mu p_\nu}{p^2}D_{\mu\nu}(p)\,, \quad
D_t(p^2) = \frac{1}{3} \Bigl(D_{\mu\mu}(p)-D_l(p^2)\Bigr)\,.
\eeqn
It follows from \eq{eq:lg-mom} that the longitudinal part of the
propagator of the diagonal gluon, $D^{diag}_l$, is zero.
Thus we have three formfactors $D^{\mathrm{diag}}_t$
and $D_{t,l}^{\mathrm{offdiag}}$.

\section{Numerical Results}
\label{sec:Numerical}

We calculate the propagators \eq{eq:diagprop}, \eq{eq:offdiagprop}
on the symmetric lattices $V=L^4$ with $L=16,24,32$ using
50, 138 and 30 configurations, respectively. Simulations are done at
$\beta = 2.40$ which corresponds to the lattice spacing
$a = (1.66 \GeV)^{-1}$ ~\cite{Ref:Beta} if one fixes the physical scale
$\sqrt{\sigma}=440$~MeV. The details of the numerical gauge fixing
procedure are given in the Appendix.

\begin{figure}[!tb]
\centering
\begin{tabular}{cc}
\includegraphics[angle=0,scale=0.3]{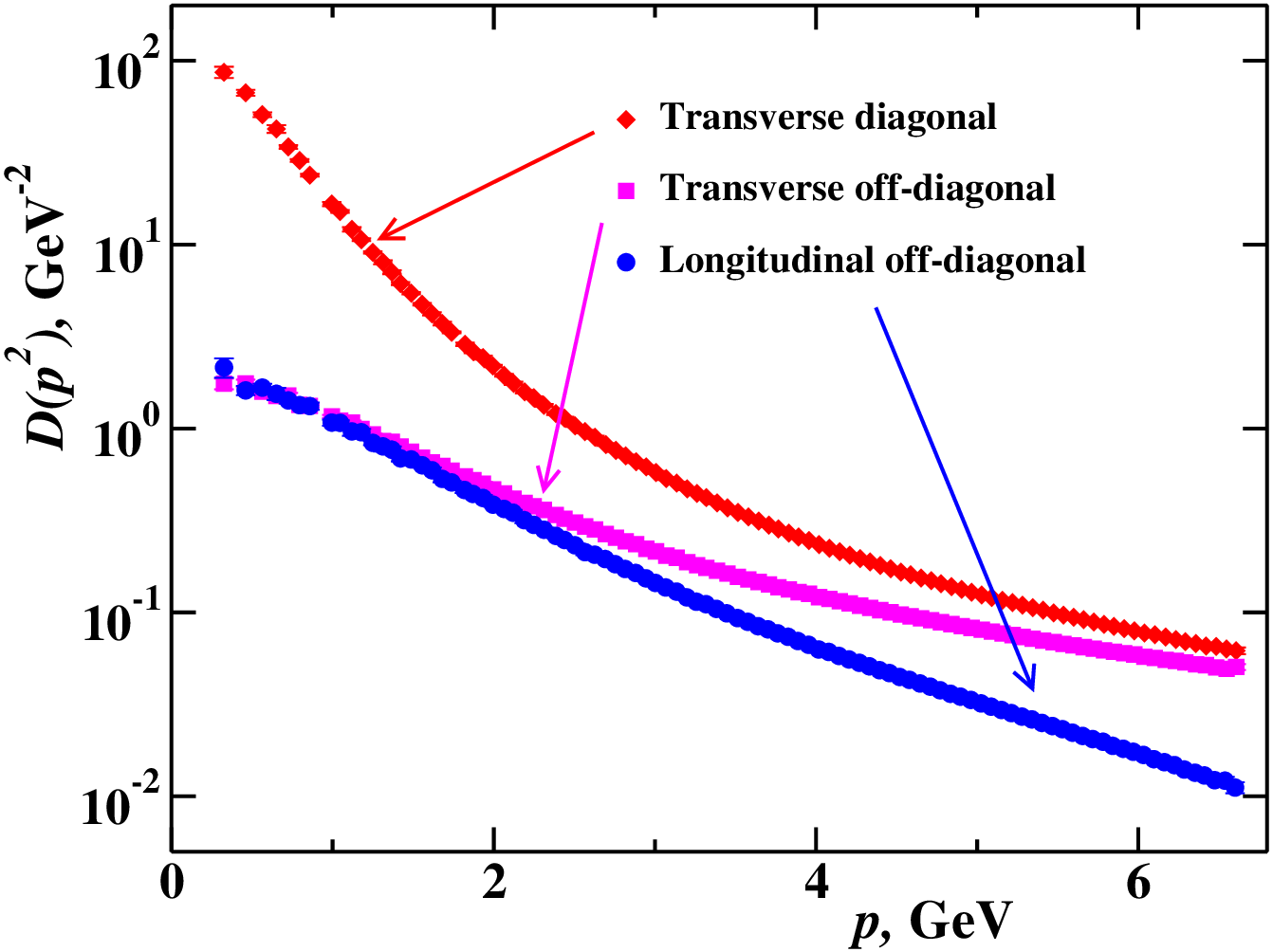} &
\includegraphics[angle=0,scale=0.3]{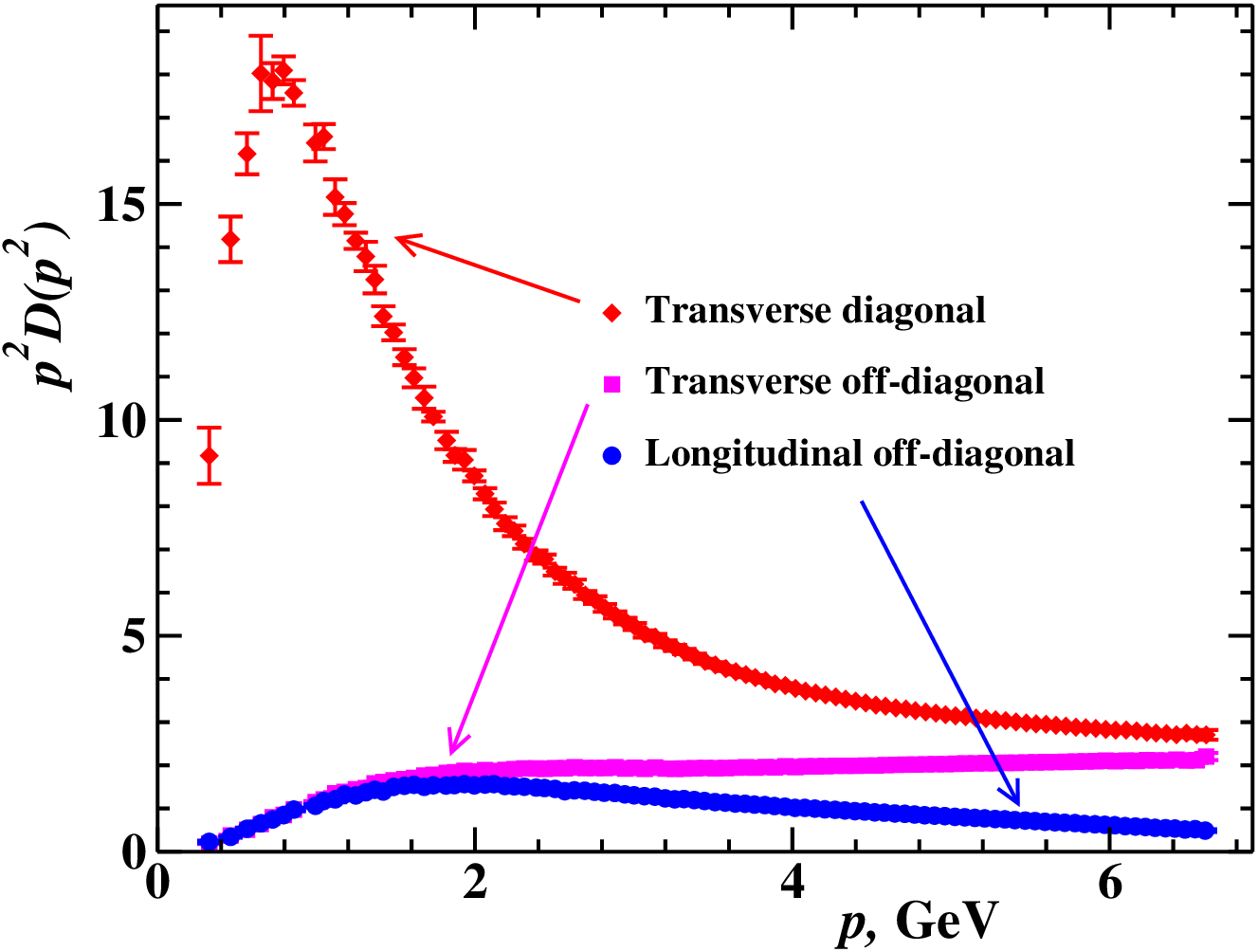} \\
(a) & (b)\\
\end{tabular}
\caption{(a) The formfactors $D(p^2)$, and (b) the gluon dressing functions,
$p^2\, D(p^2)$, {\it vs.} momentum, $p$.}
\label{fig:prop}
\end{figure}

We show all non-zero formfactors as functions of $p^2$ in Figure~\ref{fig:prop}(a)
(in all Figures of this paper we depict the data obtained on $32^4$ lattice
unless stated otherwise).
One can see that all formfactors
seem to tend to finite values as momentum goes to zero
{\it i.e.} none of them is divergent or vanishing. The diagonal
formfactor is dominating over the off--diagonal formfactors.

In Figure~\ref{fig:prop}(b) the gluon dressing function, $p^2\, D(p^2)$,
is depicted. The diagonal (transverse) dressing function
has a relatively narrow maximum at
non--zero momentum $p^{\diag}_0 \approx 0.7 \GeV$. Its behavior
at small momenta is qualitatively very similar to the behavior
of the gluon propagator in the Landau gauge (see e.g. \cite{fit12}). The
off-diagonal longitudinal dressing function has a wide maximum at
$p^{\offdiag}_0 \approx 2 \GeV$, while  for the transverse off-diagonal
dressing function formfactor it is a monotonically rising function for
all available momenta.

One can compare the propagators obtained with the $U(1)$ gauge conditions
\eq{eq:coscosfunc:U1} and \eq{eq:coscosfunc}. The first condition was implemented in
Ref.~\cite{Bornyakov:2002vv} while the last one is adopted in the present paper.
The comparison shows that the transverse diagonal propagators for these
two gauge conditions coincide with each other at large momenta. At small momenta
the formfactor obtained with Eq.~\eq{eq:coscosfunc:U1}
is slightly larger then the one calculated with Eq.~\eq{eq:coscosfunc}.
The difference at momentum $p=430\MeV$ is about 15\%.
The formfactors for the off-diagonal gluons coincide with each other
for all available momenta.

It is seen from Figure~\ref{fig:prop} that at $p \approx 6$~GeV dressing function
$p^2\, D^{diag, offdiag}_{t,l} \approx 3$ and differ
much from the free from $1/p^2$. The large renormalization of the formfactors in
Landau gauge is discussed in Ref.~\cite{Becirevic:1999uc}. There it has been shown that even
three-loop corrections do not describe the behavior of the lattice gluon
propagator at $p \approx 6$~GeV.

\begin{figure}[!tb]
%\vspace{20mm}
\centering
\begin{tabular}{cc}
\includegraphics[angle=0,scale=0.3]{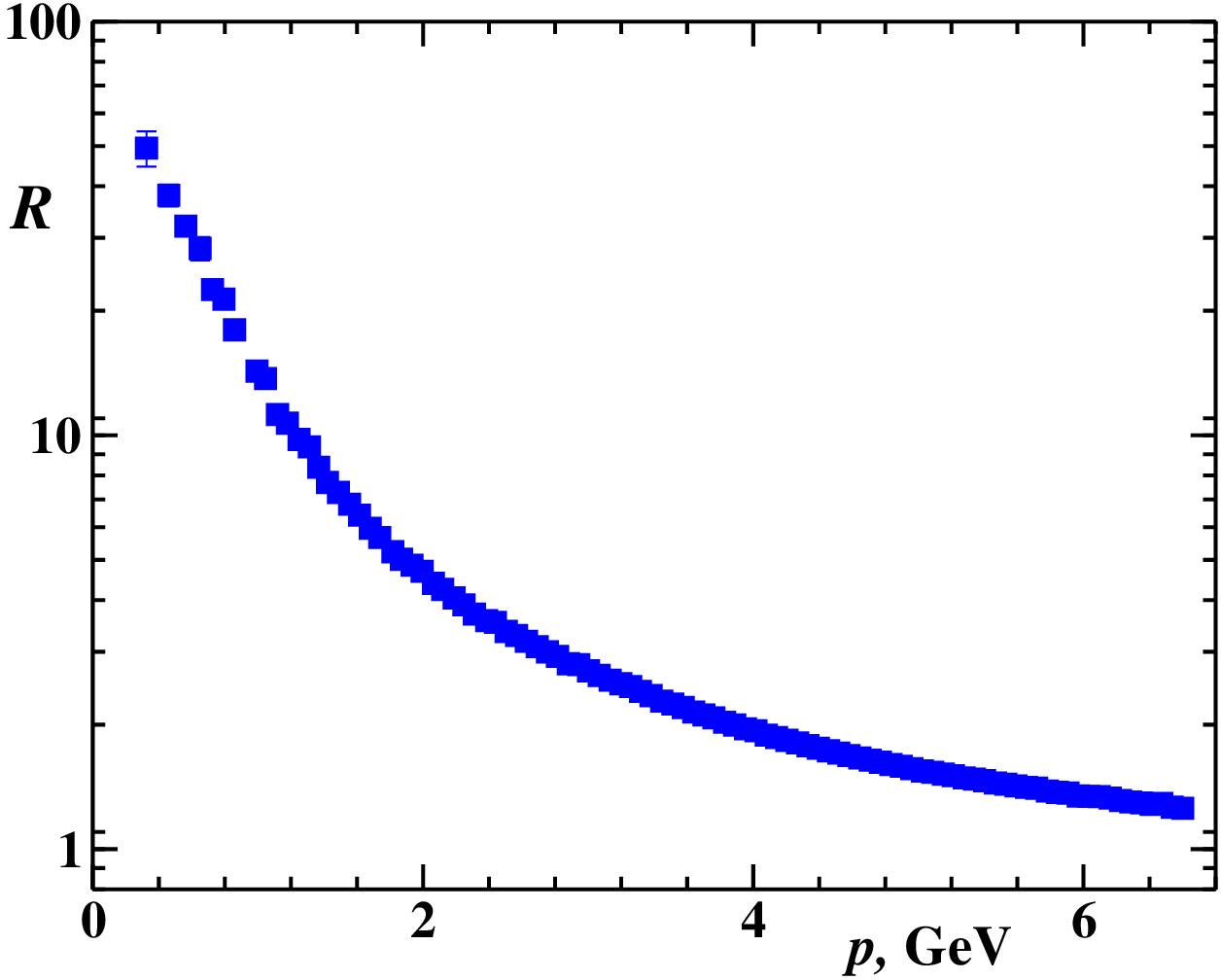} \hspace{5mm} &
\hspace{5mm} \includegraphics[angle=0,scale=0.3]{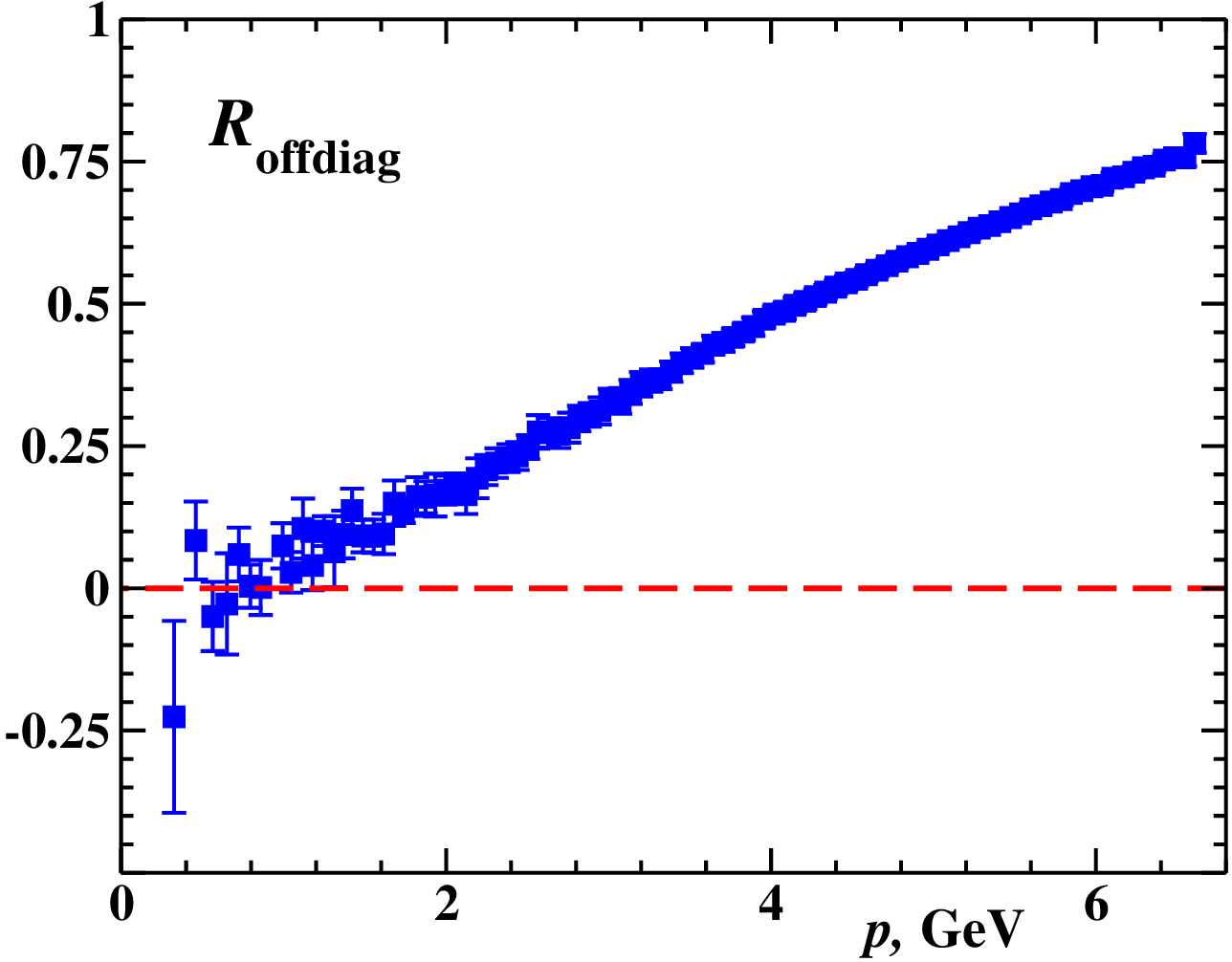}\\
(a) & (b)\\
\end{tabular}
\caption{(a) The ratio of the transverse diagonal and off--diagonal
components~\eq{eq:ratio:diag}. (b) The ratio of the off--diagonal
components~\eq{eq:ratio:offdiag} $vs.$ $p$.}
\label{fig:ratio}
\end{figure}

{}From Figures~\ref{fig:prop}(a,b) it is clear that the off-diagonal
gluon propagator is suppressed in comparison with the diagonal one.
In Figure~\ref{fig:ratio}(a) we plot the ratio
\beqn
R(p^2) = \frac{D^{\diag}_t(p^2)}{D^{\offdiag}_t(p^2)}\,.
\label{eq:ratio:diag}
\eeqn
It is seen that the suppression of the off-diagonal propagator increases as the
momentum decreases. This may be considered as an indication that
in the MA gauge the diagonal gluons are
responsible for physics in the infrared region.

{}From Figures~\ref{fig:prop}(a,b) one may also notice that the
formfactors $D^{\offdiag}_t(p^2)$ and $D^{\offdiag}_l(p^2)$ coincide
at small momenta. In Figure~\ref{fig:ratio}(b) we plot the ratio
\beqn
R_{\offdiag}(p^2) = \frac{D^{\offdiag}_t(p^2) -
D^{\offdiag}_l(p^2)}{D^{\offdiag}_t(p^2)}\,.
\label{eq:ratio:offdiag}
\eeqn
One can see that $R_{\offdiag}$ decreases with
decreasing momentum and vanishes at $p \sim 1$ GeV.
This implies that in the IR region the off--diagonal  propagator
has the form
\beqn
D^{\offdiag}_{\mu\nu}(p) \approx \delta_{\mu\nu} \cdot
D_t^{\offdiag}(p^2)\,, \quad p^2 \lesssim 1\GeV\,.
\eeqn

In order to characterize the propagators quantitatively we have
fitted the formfactors in the infrared region by the following functions:
\beqn
\label{eq:fit:MIP}
D(p^2) =  &\frac{Z\, m^{2\alpha}}{{(p^2 + m^2)}^{1+\alpha}}\,,
\qquad &\mbox{(fit 1)}\,, \\
\label{eq:fit:CIS}
D(p^2) =  & \frac{Z\, m^{2\alpha}}{p^{2(1+\alpha)} +
m^{2(1+\alpha)}}\,, \qquad &
\mbox{(fit 2)}\,,\\
\label{eq:fit:Yukawa}
D(p^2) = &\frac{Z}{p^2 + m^2}\,,\qquad & \mbox{(Yukawa fit)}\,,\\
\label{eq:fit:Yukawa2}
D(p^2) = & \frac{Z}{m^2 + p^2 + \kappa p^4/m^2}\,,\qquad &
\mbox{(Yukawa 2 fit)}\,,\\
\label{eq:fit:Gribov}
D(p^2) = & \frac{Z\, p^2}{p^4 + m^4}\,,\qquad & \mbox{(Gribov fit)}\,.
\eeqn
where $Z$, $\alpha$ , $m$ and $\kappa$ are fitting parameters.
The fitting functions~\eq{eq:fit:MIP}, ~\eq{eq:fit:CIS}, ~\eq{eq:fit:Gribov}
-- after being modified to agree at large momenta with a known perturbation
theory result -- were used to fit the gluon propagator in Landau
gauge~\cite{fit12}.
It was concluded that function~\eq{eq:fit:CIS} provided
the best fit. The fitting function~\eq{eq:fit:CIS} was also used in
the compact U(1) theory and compact Abelian Higgs model~\cite{Chernodub:2001mg}
in three dimensions.

The Yukawa fitting function, eq.~\eq{eq:fit:Yukawa}, is introduced
in order to compare
our results with results of Refs.~\cite{Amemiya:zf,Bornyakov:2002vv},
where such behavior was assumed for off-diagonal propagators.
Another interesting
possibility is to consider  the momentum dependent mass in
eq.~\eq{eq:fit:Yukawa},
$m^2 \to m^2(p^2)$. Keeping only the lowest order of $p^2$
in $m^2(p^2)$ expansion we get the fitting function~\eq{eq:fit:Yukawa2}
where $\kappa$ is an additional dimensionless parameter.
Finally, we fit the data by the function~\eq{eq:fit:Gribov}
inspired by the Gribov proposal~\cite{Gribov} for the Landau
gauge.

\begin{figure}[!tb]
\centering
\begin{tabular}{cc}
\hspace{5mm}\includegraphics[angle=0,scale=0.3]{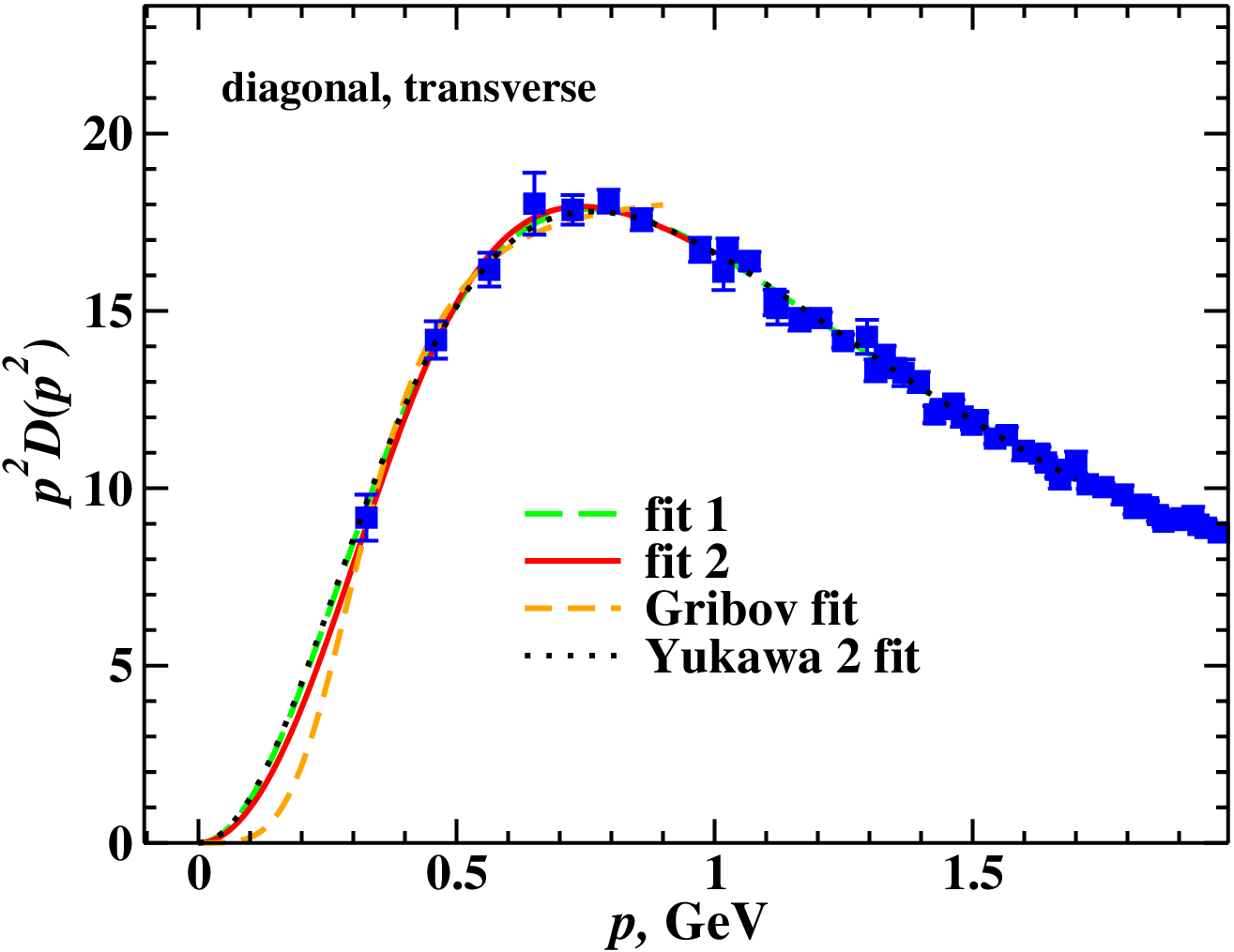}
&
\includegraphics[angle=0,scale=0.3]{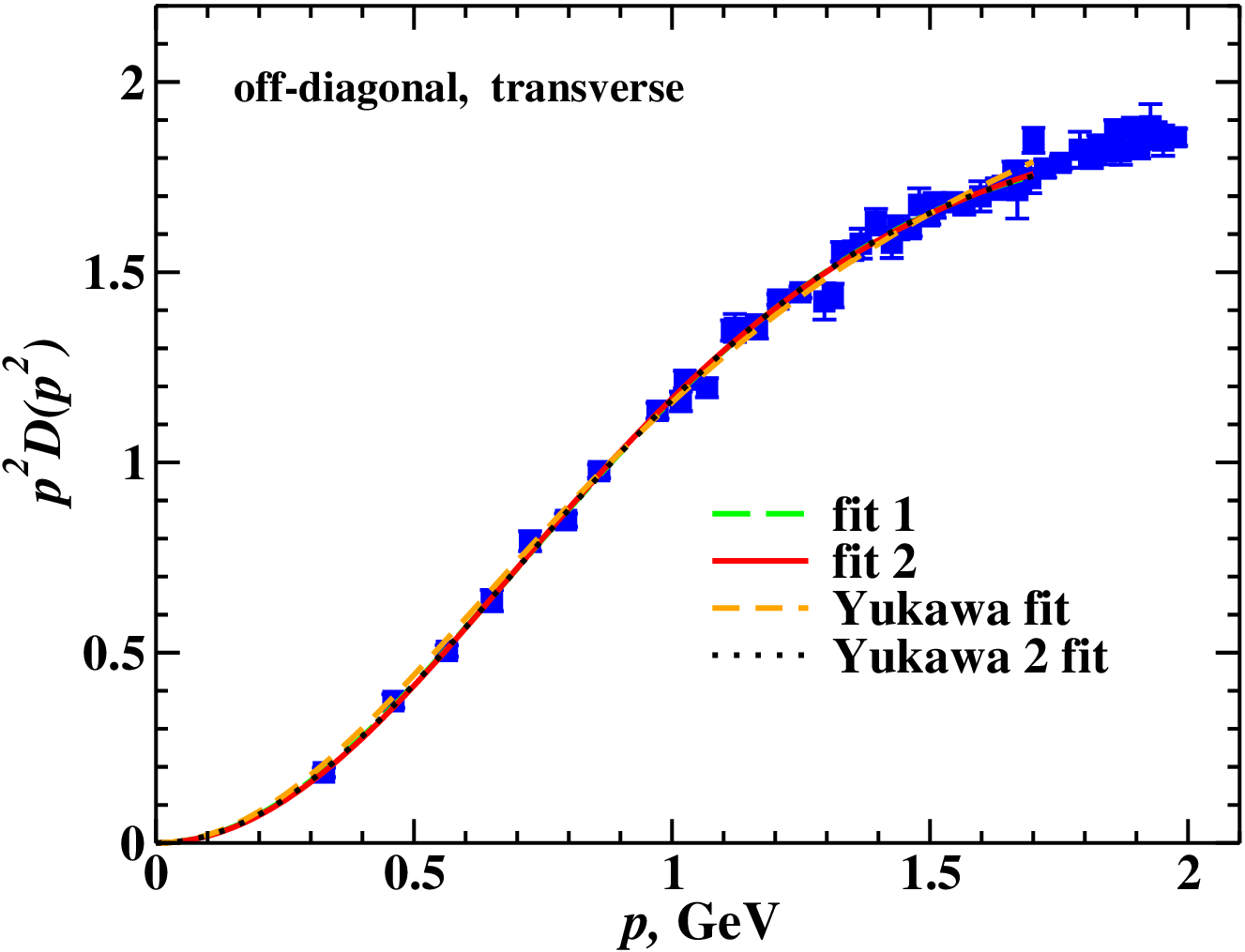} \\
~~~~~~~~(a) & ~~~~~~(b)\\
    &   \\
    &   \\
\multicolumn{2}{c}{
\hspace{5mm}\includegraphics[angle=0,scale=0.3]{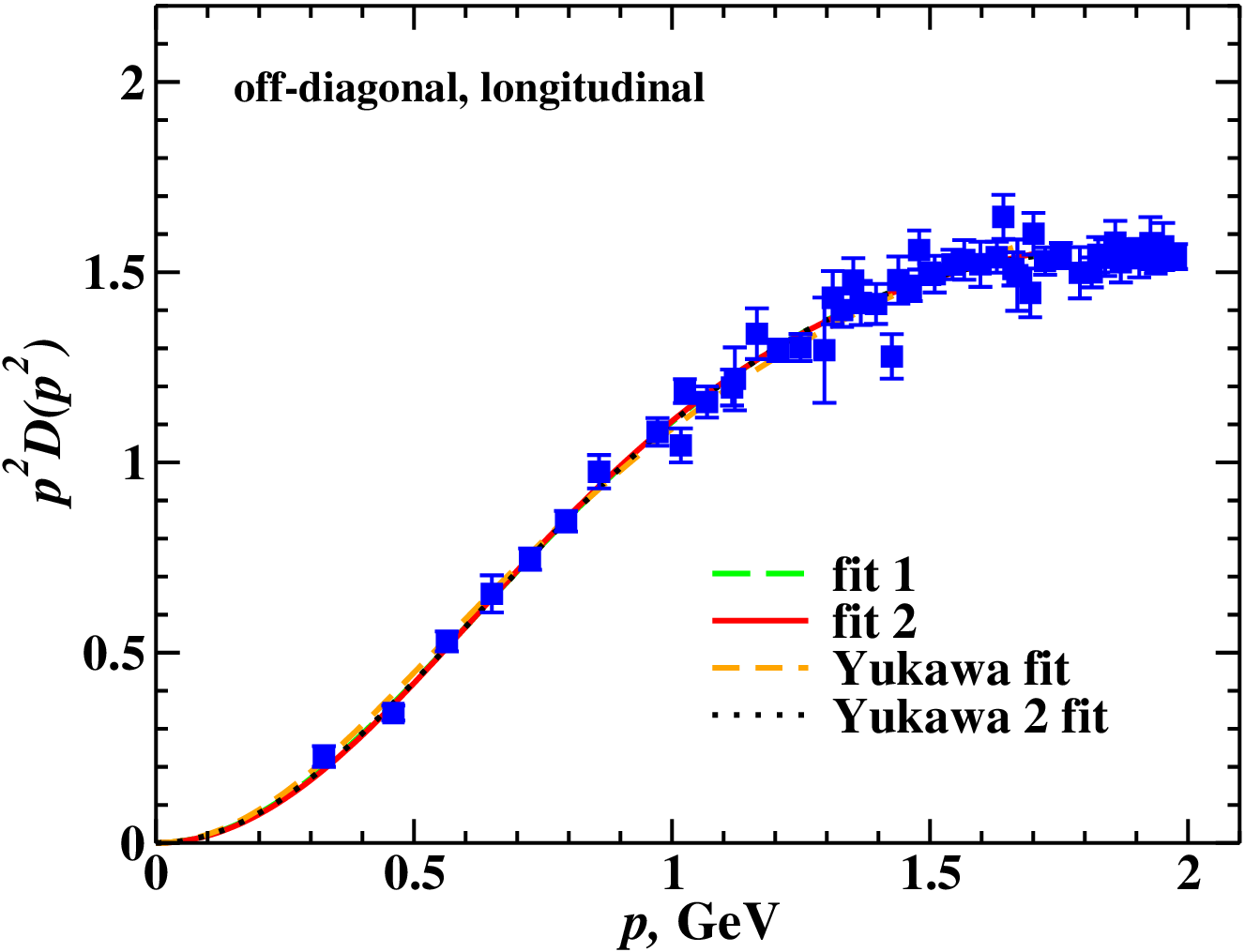}}\\
\multicolumn{2}{c}{\hspace{15mm}(c)}\\
\end{tabular}
\caption{Fits of (a) transverse part of the diagonal
propagator, (b) transverse and (c) longitudinal parts of the
off-diagonal propagators, by the functions \eq{eq:fit:MIP} and
\eq{eq:fit:CIS}.}
\label{fig:fit:prop}
\end{figure}

The quality of the fitting result depends on the interval of momentum
used in fitting. In this paper we restrict ourselves to
the infrared region. We have chosen the interval starting from
$p_{\min}= 2\pi/(32a) = 0.325\GeV$ and ending at a variable value $p_{\max}$.
For every fit we determine $p_{\max}$ as the highest momentum at which the data
points corresponding to the lowest momenta are still consistent with the given fit.
In other words, when momenta $p > p_{\max}$ are included in the
fit the fitting curves go off the error bars of the lowest momenta data
points. We employ this procedure for the diagonal propagator only,
because the statistical weight of the infrared data points is
low\footnote{Therefore the $\chi^2$--criterion can not be used for
the definition of $p_{\max}$.} while the significance of these
points is high. In the case of the off-diagonal propagators
we have limited the region of fit by highest momentum $p_{\max} =1.7\GeV$.
The fits are shown in Figures~\ref{fig:fit:prop}(a-c)
and the best fit parameters are presented in Table~\ref{table:fits}.

\begin{table}[!htb]
\begin{center}
\begin{tabular}{|l|l|l|l|c|c|}
\hline
fit  &  \hskip 2mm $m$, GeV  & \hskip 7mm $\alpha$ or $\kappa$
& \hskip 7mm Z & $p_{\max},\GeV$ & $\chi^2/dof$ \\
\hline
\multicolumn{6}{|c|}{Transverse diagonal}\\
\hline
fit 1   & 0.73(2) & 0.92(3) & 16.9(4) & 1.7  & 0.8\\
fit 2   & 0.58(2) & 0.49(5) & 8.5(2)  & 1.0  & 0.4\\
Gribov fit   & 0.33(1) & -  & 4.58(5) & 0.9  & 0.9\\
Yukawa 2 fit & 0.50(2) & 0.19(3) & 8.3(3)& 1.7 & 0.9 \\
\hline
\multicolumn{6}{|c|}{Transverse off-diagonal}\\
\hline
fit 1 & 1.6(2) & 0.6(2) & 1.3(2)  & 1.7 & 1.0  \\
fit 2 & 1.26(4) & 0.19(4) & 0.73(2) & 1.7 & 1.0  \\
Yukawa fit & 1.08(2) & 0         & 0.63(1) & 1.7 & 1.5  \\
Yukawa 2 fit & 1.29(6) &  0.15(5) & 0.81(5) & 1.7 &1.0 \\
\hline
\multicolumn{6}{|c|}{Longitudinal off-diagonal}\\
\hline
fit 1  & 1.4(2)  & 0.5(3)  & 1.0(3) & 1.7 & 1.1 \\
fit 2  & 1.14(6) & 0.19(6) & 0.63(3) & 1.7 & 1.1 \\
Yukawa fit  & 0.96(3) & 0         & 0.52(1) & 1.7 & 1.3\\
Yukawa 2 fit   & 1.14(8) & 0.12(7) &  0.66(6) & 1.7 & 1.1\\
\hline
\end{tabular}
\end{center}
\caption{The fitting results for the propagator formfactors at
low momenta: best parameters of the fits (\ref{eq:fit:MIP}--\ref{eq:fit:Gribov}).
The corresponding highest momentum, $p_{\max}$, and
$\chi^2/d.o.f.$ are also presented.}
\label{table:fits}
\end{table}

First we discuss the fits of the diagonal propagator. The three parameter
fits (\ref{eq:fit:MIP},\ref{eq:fit:CIS},\ref{eq:fit:Yukawa2}) are working well
and the corresponding curves are almost indistinguishable from each other, Figure~\ref{fig:fit:prop}(a).
However, the fitting function \eq{eq:fit:CIS} gives
twice smaller value of $p_{\max}$ than the other functions, indicating that fit
\eq{eq:fit:CIS} works in a narrower region than the other fits. The mass parameters
for these fits do not coincide, and the difference between them is about $30\%$.

We have also applied the two-parameter fits given by Yukawa \eq{eq:fit:Yukawa} and
Gribov \eq{eq:fit:Gribov} functions. The Gribov fit is working well for the
diagonal propagator. One can see from Figure~\ref{fig:fit:prop}(a) that
in order to discriminate the Gribov fitting function from the others we need
the data at momenta smaller than available in our study.
The Yukawa fit of the diagonal propagator does not work at all (we get
$\chi^2 / d.o.f. \sim 6$ for fits in $p_{\max} < 1 \GeV$ region).

Concerning the transverse and longitudinal parts of the off-diagonal
propagator one can make a few observations.
First we notice, that the Gribov fitting function~\eq{eq:fit:Gribov} is clearly
not applicable for fitting of these propagators. Second, one can see that
the formfactors for transverse and longitudinal parts
almost coincide with each other at small momenta. The last fact implies that the best fit parameters
for each particular type of the fits (\ref{eq:fit:MIP}-\ref{eq:fit:Yukawa2}) must
coincide as well,
$$
m^{\offdiag}_{t} \approx m^{\offdiag}_{l}\,,\quad
Z^{\offdiag}_{t} \approx Z^{\offdiag}_l\,, \quad
\alpha^{\offdiag}_{t} \approx \alpha^{\offdiag}_l\,, \quad
\kappa^{\offdiag}_{t} \approx \kappa^{\offdiag}_l\,,
$$
in agreement with Table~(\ref{table:fits}).

We have also found that the mass parameters in the
off-diagonal gluon fits are approximately two times bigger then the
corresponding parameters in the diagonal fit propagators:
$$
m^{\offdiag}_{t,l} \approx 2 m^{\diag}_t\,.
$$
Thus, the off-diagonal propagator is clearly short--ranged compared
to the diagonal one.

In Ref.~\cite{Amemiya:zf} the off-diagonal propagator was successfully fitted in
the infrared region of {\it coordinate space} by the Yukawa propagator. Our
results (with respect to the off--diagonal propagator) indicate, that other
fitting functions can also be used to describe this propagator.

Summarizing, we can say that the diagonal propagator can be
fitted almost equally well with any of the  three-parameter fits
(\ref{eq:fit:MIP}), (\ref{eq:fit:CIS}) and (\ref{eq:fit:Yukawa2}). The same
is true for the off-diagonal propagator. Among the two-parameter fits
(\ref{eq:fit:Gribov}) is superior for the diagonal propagator while
(\ref{eq:fit:Yukawa}) is better for the off-diagonal one.

At smallest available momentum (325~MeV) the off-diagonal propagator
is suppressed by the factor about 50 with respect to the diagonal one.
Note that this suppression is only partially due to larger values of the mass
parameter $m$ while equally or even more important role is played by smaller
values of the parameter $Z$. We expect that the similar suppression exists also
in the continuum theory because at our lattice spacing the renormalization
effects for the $Z$--parameter are already quite small (according to
Figure~\ref{fig:prop}(a) the transverse diagonal and off--diagonal formfactors
almost coincide with each other at largest available momentum).

\section{Conclusions}

Our results obtained in the Maximally Abelian gauge of $SU(2)$
gluodynamics clearly show that at low momenta the propagator of
the diagonal gluon is much larger than the propagator of  the
off-diagonal gluons. This suggests that the colored objects at large
distances interact mainly due to exchange by the diagonal gluons
in agreement with the Abelian dominance property established in numerical
studies of the MA gauge~\cite{Suzuki:1989gp} for fundamental test
charges\footnote{As for adjoint charges see discussion in
Ref.~\cite{SuzukiChernodub2002}.}.

The propagators do not show indications that they are either vanishing or
divergent when $p \rightarrow 0$. To provide a quantitative description of the
propagators at low momenta we fit the propagator formfactors using various
functions. All infrared fits for both diagonal and off-diagonal
propagators contain massive parameters, which are non-zero
for both propagators. When the same  fitting function is applied to the diagonal
and off-diagonal formfactors the mass parameter for off-diagonal
formfactor is more than twice bigger than that for the diagonal
one. This is in a qualitative agreement with findings of Ref.~\cite{Amemiya:zf}.
But our more detailed analysis revealed that the difference in
values of the mass parameters is not the only reason of the off-diagonal
propagator suppression, the other reason is small values of parameter~$Z$.

We have found that the diagonal propagator has qualitatively the same
momentum dependence as the gluon propagator in Landau gauge while
the off-diagonal propagator is very different.
At small momenta the off-diagonal propagator is
diagonal with respect to the space indices and thus defined by a single scalar function
since its transverse and longitudinal formfactors become equal within error bars.

\section*{Acknowledgments}

The authors are grateful to A.~Schiller, E.--M. Ilgenfritz, G.~ Schierholz
and V.I.~Zakharov for interesting discussions, and to K.--I. Kondo for
useful suggestions. M.~I.~P is partially supported by grants RFBR
02-02-17308, RFBR 01-02-17456, RFBR 00-15-96-786, INTAS-00-00111, and CRDF
award RPI-2364-MO-02. S.~M. is partially supported by grants RFBR 02-02-17308
and CRDF award MO-011-0. F.~V.~G. is supported by grant RFBR 03-02-16-941.
M.~N.~Ch. is supported by the JSPS Fellowship P01023.

\section*{Appendix}
\setcounter{equation}{0}
\renewcommand\theequation{A\arabic{equation}}

The Maximally Abelian and the Abelian Landau gauges are defined as
global maxima of the functionals \eq{eq:MAG:global} and
\eq{eq:coscosfunc}, respectively. The global maxima is
difficult to reach numerically and usually one finds several
field configurations corresponding to  local maxima of the gauge fixing
functional and then the configuration with the highest
value of the functional is chosen. The choice of the correct
maximum, known as a Gribov problem~\cite{Gribov}, is
crucial for the observables both in the MA gauge of $SU(2)$ gauge
model~\cite{Bali:1996dm} and in the Landau gauge of the U(1) gauge
model~\cite{Nakamura:ww}.

Gauge fixing of the MA gauge with a careful treatment of the Gribov
ambiguity was studied in Ref.~\cite{Bali:1996dm}. In our
investigation we follow this paper using the Simulated Annealing
algorithm with 10 randomly generated gauge copies.
The MA gauge fixing procedure is described in~\cite{Bali:1996dm},
and the Abelian Landau gauge fixing algorithm is briefly considered below.

To maximize the functional~(\ref{eq:coscosfunc}) we use the local
over-relaxation algorithm with $\omega = 1.8$, see, {\it e.g.},
Ref.~\cite{Cucchieri:1995pn}, with 20 randomly generated
gauge copies. For the local gauge fixing
we choose the following convergence criterion:
\beq
\label{eq:stop2}
{\cal G}(x) = |\sum_{\mu=1}^4 \cos\varphi_\mu(x)\sin\theta_\mu(x)
              - \cos\varphi_\mu(x+\mu)\sin\theta_\mu(x+\mu)| \leq
              \varepsilon\,,
\eeq
where $\varepsilon$ is a small parameter. Note, that this condition
must be satisfied at each site $x$ of the lattice.

\begin{figure}[!tb]
\centering
\begin{tabular}{cc}
\includegraphics[angle=0,scale=0.3]{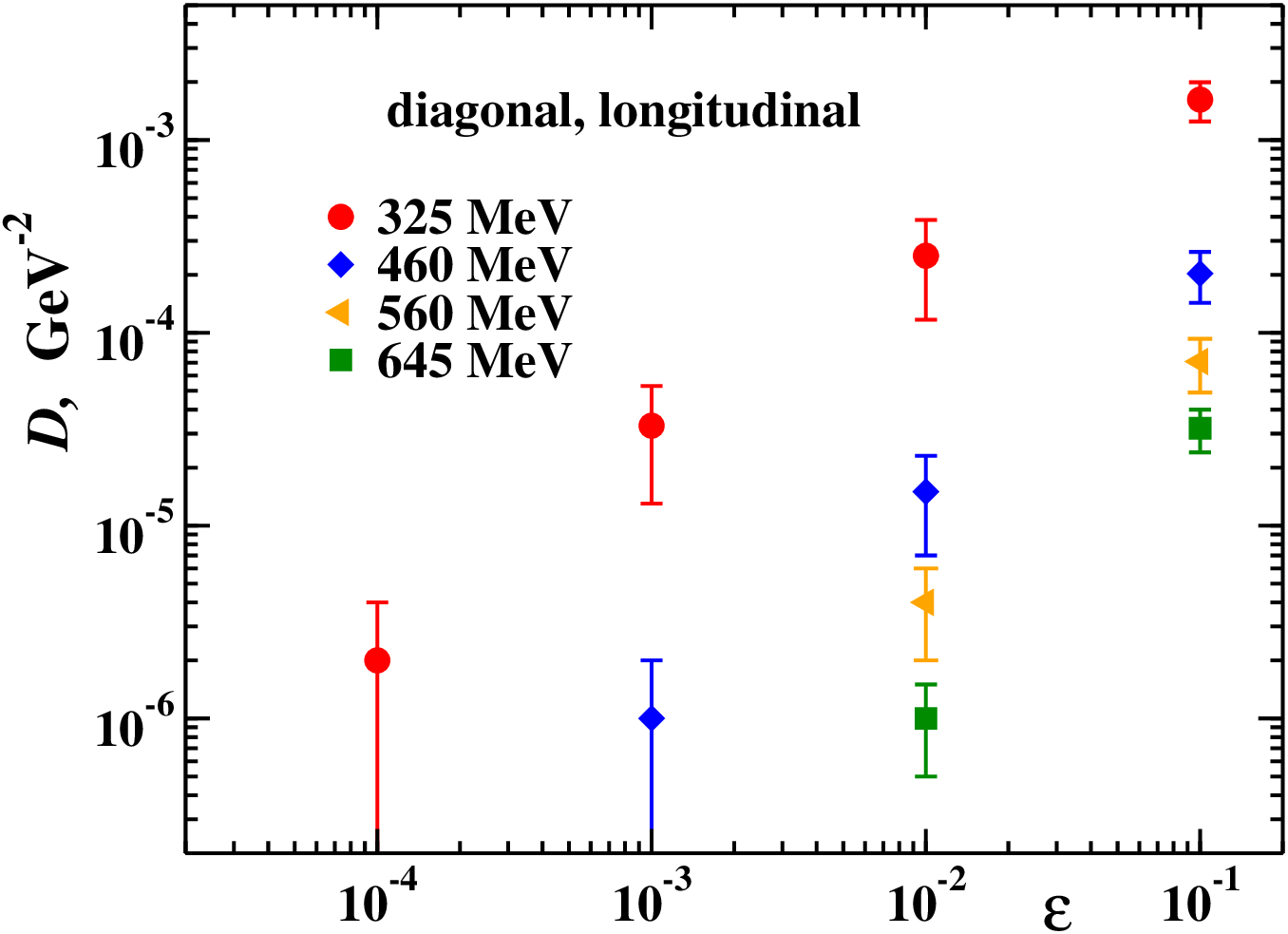} \hspace{5mm}& \hspace{5mm}
\includegraphics[angle=0,scale=0.3]{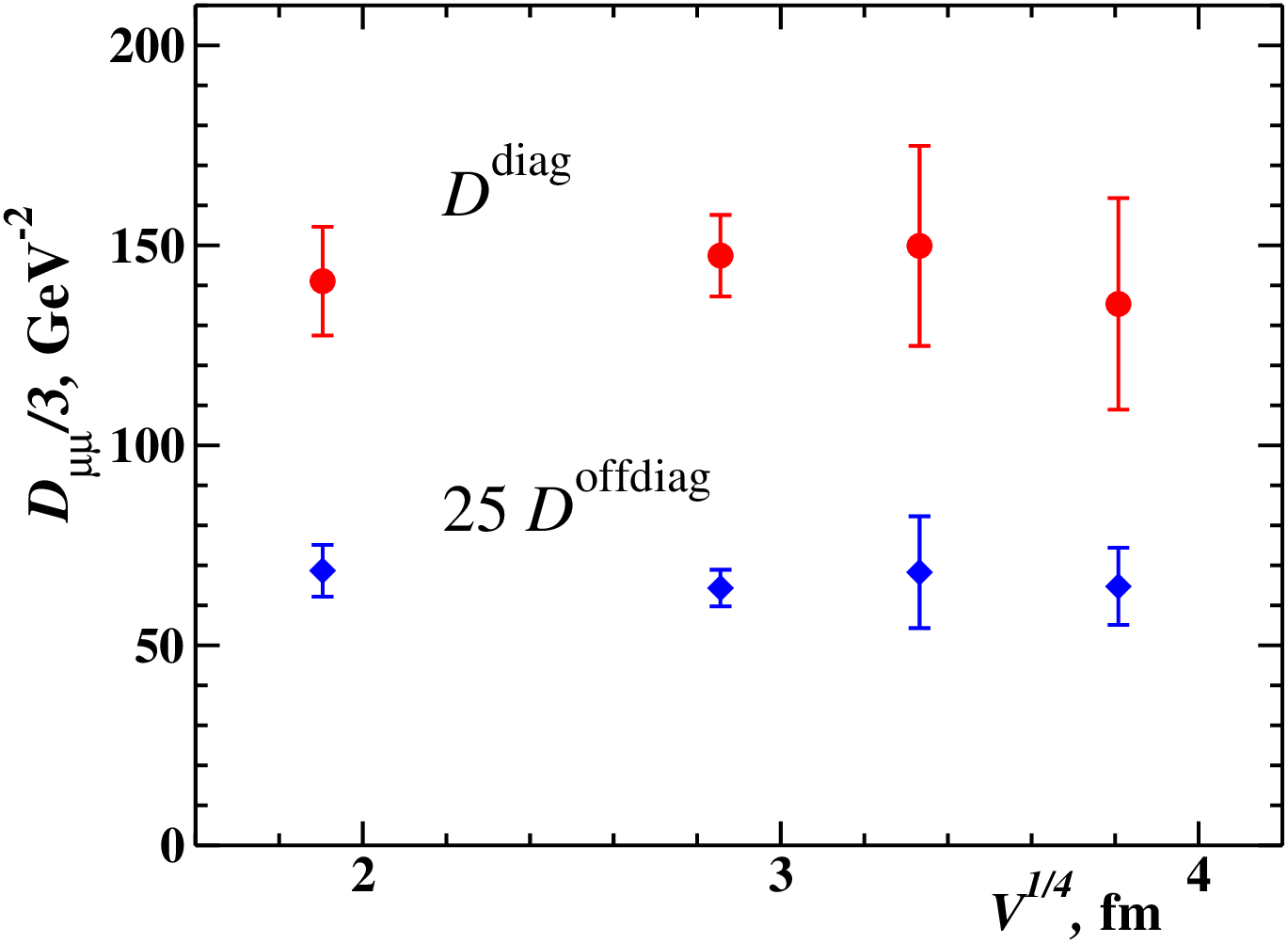}
\end{tabular}
\caption{(a) The longitudinal part of the propagator of the
diagonal gluon at various momenta $vs.$ convergence parameter
$\varepsilon$; (b) the volume dependence of the diagonal and
off-diagonal propagators at $p=0$, the corresponding lattice volumes are:
$16^4$, $24^4$, $28^4$ and $32^4$.
}
\label{fig:epsdep}
\end{figure}

To study the effects of the incomplete gauge fixing we chose the longitudinal
part of the propagator of the diagonal gluon $D^{\mathrm{diag}}_l$
because it is most sensitive to the details of the gauge fixing procedure.
According to the local gauge condition~\eq{eq:contland}, $D^{\mathrm{diag}}_l$
must be zero when perfect numerical procedure is used. Its dependence on the
convergence parameter $\varepsilon$ is presented in Figure~\ref{fig:epsdep}(a).
The propagator significantly depends on $\varepsilon$, especially in
the region of small momenta. In our simulations we choose $\varepsilon=10^{-6}$.

We have also checked the dependence of $D^{\mathrm{diag}}_t$
on the number of gauge copies, $(N_{gc})$, used
both in the MA  and in the Abelian Landau gauge fixings. This check has been
done using  set of 30 configurations on $L=24$ lattice. Within
error bars we have observed no dependence on the number of
the MA gauge  Gribov copies and we have found only  very mild dependence on
the number of the Abelian Landau gauge copies.

To check the finite volume effects
we have calculated the propagators on
different lattices at the same $\beta = 2.40$.
The transverse formfactors at zero momentum $D^{diag, offdiag}_t(0)$ were
calculated as follows:
\beq
\label{eq:zmformfactors}
D^{diag, offdiag}_t(0) = \frac{1}{3}D^{diag, offdiag}_{\mu\mu}(0).
\eeq
We find that
within error bars the values of the zero-momentum propagator
are independent of the lattice volume, as can be seen from
Figure~\ref{fig:epsdep}(b).

\begin{figure}[!tb]
\centering
\includegraphics[angle=0,scale=0.3]{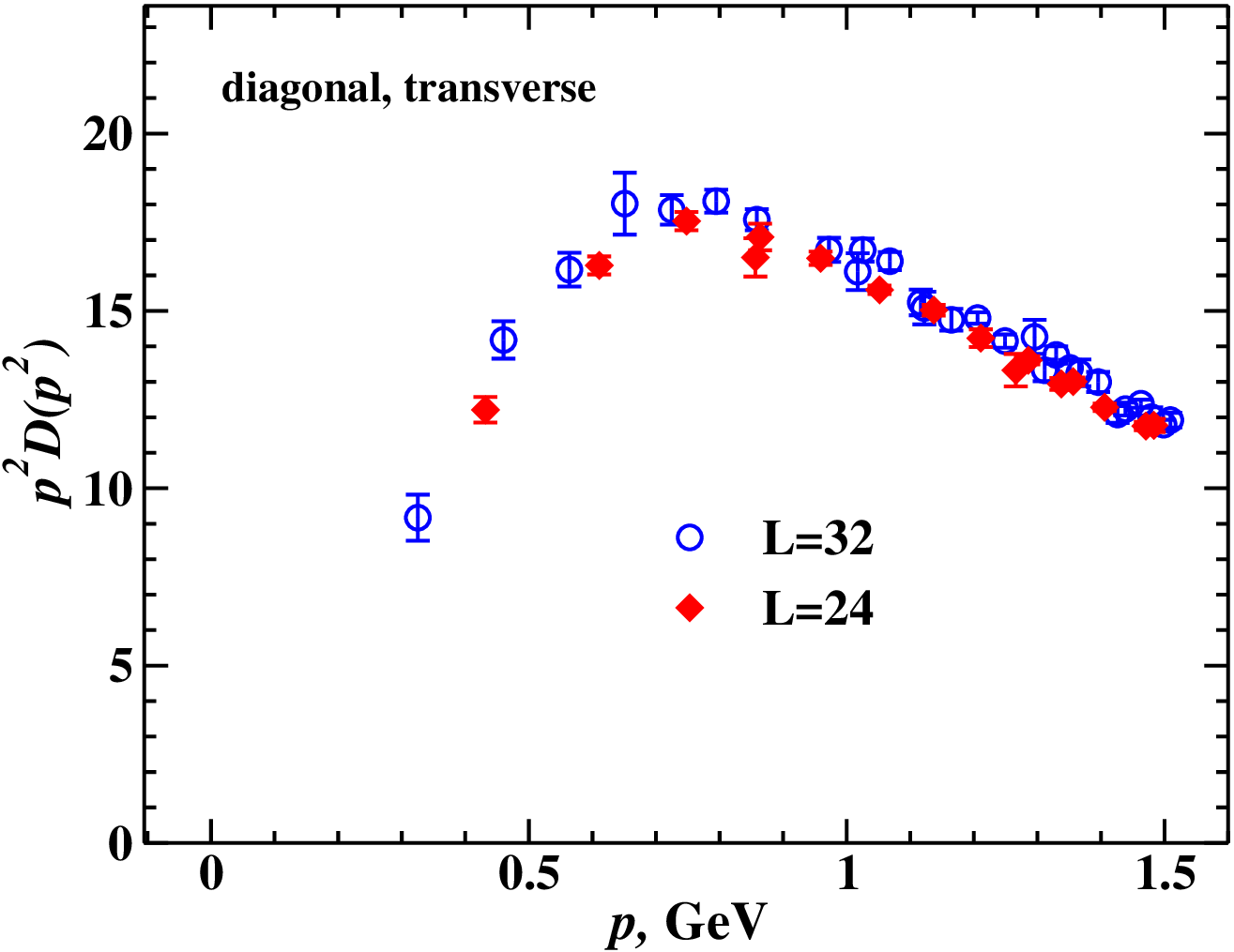}
\caption{The transverse diagonal propagator at small momenta for the
lattices $L=24$ and $L=32$.}
\label{fig:smallmomentum}
\end{figure}

We have also studied the volume dependence of the transverse part
of the diagonal propagator at non--zero momenta. Since the finite
volume affects mainly the low momentum region we have concentrated on
the propagator at $p<1.5$~GeV. In Figure~\ref{fig:smallmomentum} we
plot the transverse part of the diagonal propagator calculated on
$24^4$ and $32^4$ lattices. It is seen that the volume dependence is
very weak, it is within the error bars. The position of the maximum of
$p^2 D(p^2)$ is the same for both lattices. The fit of the data for
the propagator on $24^4$ lattice gives parameters which differs less
than by $4\%$ from that given in Table~1 for $32^4$ lattice. Thus we
estimate the systematic error induced by the finite volume effects to
be less than $4\%$.

\end{document}